\documentclass[]{jpconf}
\usepackage{graphicx}
\usepackage{dcolumn}
\usepackage{bm}
\newcommand{\aF}{$\alpha^{2}F(\omega)$ }
\newcommand{\glue}{$\tilde{\Pi}(\omega)$ }
\newcommand{\glueMFL}{$\tilde{\Pi}_{MFL}(\omega)$ }
\newcommand{\glueSF}{$\tilde{\Pi}_{SF}(\omega)$ }
\newcommand{\gluehg}{$\tilde{\Pi}_{HG}(\omega)$ }

\graphicspath{{graphs/}}
\begin{document}



\title{Optics clues to pairing glues in high T$_{c}$ cuprates}

\author{E. van Heumen, A.B. Kuzmenko, D. van der Marel}

\address{D\'epartement de Physique de la Mati\`ere Condens\'ee,
Universit\'e de Gen\`eve, quai Ernest-Ansermet 24, CH1211 ,
Gen\`eve 4, Switzerland}

\ead{e.vanheumen@uva.nl}

\begin{abstract}
We analyze optical spectra of the high temperature
superconductor HgBa$_{2}$CuO$_{4+\delta}$ using a minimal model
of electrons coupled to bosons. We consider the marginal Fermi
liquid theory and the spin fluctuation theory, as well as a
histogram representation of the bosonic spectral density. We
find that the two theories can both be used to describe the
experimental data provided that we allow for an additional
scattering channel with an energy of 55 meV.

\end{abstract}



\section{Introduction}
\begin{figure}[bht]
\begin{minipage}{17pc}
\includegraphics[width=17pc]{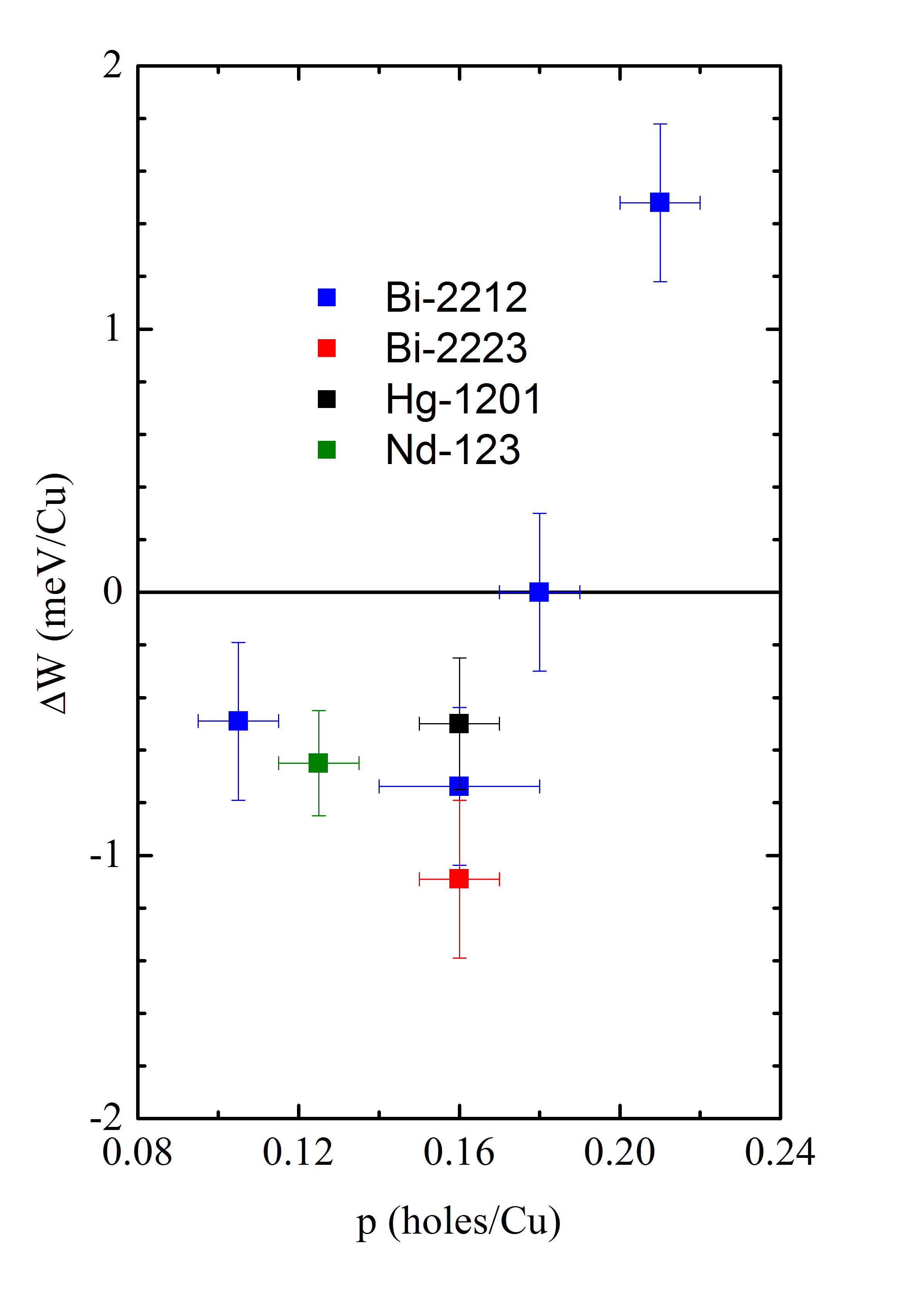}
\end{minipage}\hspace{2pc}%
\begin{minipage}[b]{18pc}
\caption{\label{dopdepSW} Doping dependence of the spectral weight change $\Delta W$ at T$_{c}$.
The numbers are taken from Ref.
\cite{molegraaf-science-2002,carbone-PRB-2006a} for Bi-2212,
from Ref. \cite{carbone-PRB-2006b} for Bi-2223 and from Ref.
\cite{heumen-PRB-2007} for Hg-1201. The result for Nd-123 is
unpublished. This result is similar to the one obtained by the group of Bontemps \cite{deutscher-PRB-2005}.}
\end{minipage}
\end{figure}

The puzzle of superconductivity in the cuprates remains
unsolved. Over the years it has become clear that the nature of
the normal state is an equally important puzzle to solve. It
now seems established that the strongly overdoped cuprates
posses a well defined Fermi surface
\cite{hussey-nat-2003,damascelli-RMP-2003,takeuchi-PRL-2005,plate-PRL-2005}
indicating that at least in this part of the phase diagram at
low energy a Fermi liquid picture might apply for the normal
state. The underdoped cuprates are characterized by so-called
Fermi arcs: incomplete pieces of Fermi surface
\cite{norman-nat-1998}. One might expect that in this part of
the phase diagram the Fermi liquid paradigm is no longer
appropriate and the physics is indeed described by that of a
lightly doped anti-ferromagnetic insulator. Recent de Haas -
van Alphen experiments however seem to indicate that the arcs
should be interpreted in terms of small pockets
\cite{leyraud-nat-2007} thus placing the state underlying the
pseudogap state also in the Fermi liquid regime. Regardless the
nature of the superconducting and normal state, the transition
from one to the other is driven by the minimization of the free
energy, which consists of potential ($E_{pot}$) and kinetic
energy ($E_{kin}$). In conventional superconductors the
transition is driven by a lowering of the potential energy
while at the same time the formation of the Cooper pairs goes
at the cost of a smaller amount of kinetic energy. Several
experiments
\cite{basov-science-1999,molegraaf-science-2002,syro-EPL-2003}
have found that this situation might be different in the
cuprates. In \cite{basov-science-1999} the experimental results
could be explained if it was assumed that superconductivity is
driven by a gain in the $c$-axis $E_{kin}$ while in the latter
two cases the result can be interpreted in terms of a gain in
$ab$-plane $E_{kin}$. In both cases it was argued that the
different sign of $E_{kin}$ with respect to conventional
superconductors arises from the properties of the normal state
(see i.e. Ref. \cite{ioffe-science-1999} for the first case and
Ref. \cite{norman-PRB-2000,dirk-proc-2003} for the second). In
later experiments
\cite{carbone-PRB-2006a,carbone-PRB-2006b,heumen-PRB-2007} the
dependence of the change in $E_{kin}$ on doping and the number
of layers was studied. The results of these studies are
summarized in figure \ref{dopdepSW}. The quantity $\Delta W$ is
a measure of the kinetic energy change at the superconducting
transition as defined in \cite{kuzmenko-PRB-2005}.
Interestingly, $\Delta W$ changes sign for slightly overdoped
samples. Assuming that the mechanism of superconductivity does
not depend on doping, the sign change must arise from a change
in the properties of the normal state.

A possible interpretation of this result is that the
superconducting dome covers a quantum critical point
\cite{tallon-physc-2001,varma-PRL-2007}. In this scenario the
underdoped and overdoped materials would be distinct phases
explaining the change in Fermi surface topology and the change
in sign of the superconductivity induced spectral weight. It
has also been suggested that the optical properties of
optimally doped Bi-2212 \cite{dirk-nat-2003} and Bi-2223
\cite{dirk-annphys-2006} can be understood from quantum
critical fluctuations. In \cite{dirk-nat-2003} it was pointed
out that at low energy the optical spectra could be described
as a universal function of $\omega/T$ while in the mid-infrared
range the conductivity could be described by a powerlaw
instead. However, as pointed out in \cite{dirk-annphys-2006}
these two observations cannot be made consistent with each
other without assuming a non-universal background to the
optical conductivity.

Norman and Chubukov \cite{norman-PRB-2006} showed that the
powerlaw behavior could also have another interpretation: it
follows that the optical conductivity has an approximate
powerlaw behavior if one assumes a model in which electrons
interact with a broad bosonic spectrum ($\tilde{\Pi}(\omega)$).
Using the same model the anomalously large temperature
dependence of the normal state spectral weight could also be
explained \cite{norman-PRB-2007} but not the observed change in
kinetic energy at T$_{c}$ \cite{marsiglio-PRB-2008}. Stimulated
by these positive results we analyzed the optical spectra of 10
different compounds using a histogram representation for the
bosonic spectrum \cite{heumen-sub-2008} and found that this
spectrum strongly depends on doping and temperature. The doping
dependence of \glue might give an explanation for the very
different Fermi surfaces observed by ARPES for under- and
overdoped cuprates, while the temperature dependence of the
spectra could give an explanation for the (approximate)
$\omega/T$ scaling observed in \cite{dirk-annphys-2006}. Here
we will discuss two possible models for the normal state and
compare it to the earlier obtained histogram representation.

\section{Methods}\label{methods}
In normal metals the electron-phonon (EP) interaction can be
described in a framework where fermionic quasiparticles, the
electrons, interact with a spectrum of bosonic modes, the
phonons. The function describing the energy dependent coupling
is often indicated as $\alpha^{2}F(\omega)$. A closely related
theory is the Eliashberg theory of superconductivity describing
the occurrence of superconductivity in normal metals. It is an
improvement of the BCS theory of superconductivity because it
takes into account the retardation of the EP interaction. \aF
also appears in this theory and is the main determining factor
of the critical temperature at which superconductivity occurs.
It is also possible to consider bosons other than phonons, for
example plasmons, magnons or excitons. An important distinction
between these bosons and phonons is that they are made up out
of the same degrees of freedom which form the Cooper pairs: the
electrons themselves.

We compare different models for the electron-boson coupling
function, indicated here by the symbol $\tilde{\Pi}(\omega)$.
The calculation of the optical conductivity for given \glue is
straightforward (see \ref{App}). The inverse problem of
extracting \glue from experimental data can be done, but with
limited accuracy due to the progression of errors due to
experimental noise \cite{marsiglio-PLA-1998,dordevic-PRB-2005}.
Several theoretical models for the cuprates predict that the
electrons interact with a bosonic spectrum.

\subsection{Marginal Fermi liquid.}
The Marginal Fermi Liquid (MFL) theory
\cite{varma-PRL-1989,varma-PRL-2007} predicts that \glue is
given by,
\begin{equation}\label{MFL}
\tilde{\Pi}_{MFL}(\omega ) = \Lambda \tanh (\frac{\omega }{{2T}})f(\omega,\omega_c ,\Delta )
\end{equation}
where $\Lambda$ is an overall coupling constant, $T$ is the
temperature and $f(\omega,\omega_c,\Delta)$ is a high energy
cutoff function the precise shape of which is unimportant. We
have used
$f(\omega,\omega_c,\Delta)=1/(1+exp(\frac{(\omega-\omega_{c})}{\Delta}))$
as proposed originally by Varma \cite{varma-PRL-1989}. An
important feature of this spectrum is that it only depends on
$\omega/T$. It is easy to see using Eq.'s
\ref{Kubo}-\ref{Lfunc} that in this case one retrieves the
linear temperature dependence of the resistivity seen in many
experiments. It would also give rise to a conductivity that
scales as $\sigma(\omega,T)=T^{-1}g(\omega/T)$ as proposed in
\cite{dirk-annphys-2006}.

\subsection{Spin fluctuation theory.}
Several other theoretical models are based around the notion
that anti-ferromagnetic fluctuations play an important role in
understanding the physics of the (underdoped) cuprates
\cite{scalapino-PRB-1986,millis-PRB-1990,chubukov-EPL-1997}. In
these models one uses the imaginary part of the dynamic spin
susceptibility $\chi''(\omega,\vec{q}=0)$ as a measure for the
bosonic spectral density which in the ungapped state is given
by,
\begin{equation}\label{SF}
\tilde{\Pi}_{SF}(\omega ) = \frac{\Gamma\omega}{\gamma^{2}  + \omega^{2}}\quad\omega  \le
\omega _c
\end{equation}
Here $\Gamma$ is a coupling strength and $\omega_{c}$ a cutoff.
The spectrum has a maximum determined by $\omega = \gamma$. We
refer to this model as the spin fluctuation (SF) model. Note
that this form is valid only if there is no (pseudo)gap.
Although the temperature dependence is not explicitly mentioned
in Eq. \ref{SF} the parameters $\gamma$ and $\Gamma$ depend on
temperature. \glueSF does not scale as $\omega/T$ unless both
$\gamma$ and $\Gamma$ are depending linear on temperature.  The
temperature dependence of the MFL and SF models are an
important point of contrast with the EP interaction which leads
to a temperature independent $\alpha^{2}F(\omega)$. The fact
that most experimental studies find that for the cuprates \glue
has to be made temperature dependent is one of the arguments
against the interpretation of \glue in terms of phonons.

\subsection{MFL and SF + phonon models}
Finally we allow for a Lorentzian oscillator in addition to the
MFL and SF model given by,
\begin{equation}\label{Lor}
\tilde{\Pi}_{Lor}(\omega) = \frac{f_{1}^{2}\Gamma_{1}\omega}{(\omega\Gamma_{1}) ^2  + (\omega_{1}^{2}-\omega^2)^{2}}
\end{equation}
Here $f_{1}$ is the oscillator strength, $\Gamma_{1}$ is the
width and $\omega_{1}$ is the center frequency. Such a peak can
be used to describe the coupling of electrons to phonons or to
the spin resonance.

\subsection{Histogram representation.}
Although we cannot directly invert the experimental data to
obtain \glue we can use a Levenberg-Marquardt optimization
routine to determine the important features in \glue. We do
this by making a histogram of \glue using $i$ blocks with
flexible widths and heights with a total of $2i$ parameters,
\begin{equation}\label{HG}
\tilde{\Pi}_{HG}(\omega)=f_{i}\quad\omega_{i-1}\le\omega\le\omega_{i}
\end{equation}
where $\omega_{0}$ = 0 and $f_{i}$ is the height of the $i$th
block. We find that i=6 is the optimal number of blocks giving
sufficient detail to \glue without overdetermining it. For
$\omega_{0}\le\omega\le\omega_{1}$ we use
$\Pi_{HG}(\omega)=f_{1}\omega$ to circumvent the divergence of
the integral in Eq. \ref{SE}. The resulting histogram allows us
to extract a rough impression of \glue and give an indication
of the important features in the spectrum. This model function
allows us to optimize \glue such that adding more detail does
not significantly improve the result. This model was used in a
recent study of 10 different compounds \cite{heumen-sub-2008}.
We showed that \gluehg has three important features:
\begin{itemize}
\item \gluehg has to be temperature dependent.
\item \gluehg extends to very high energies (about 400 meV)
    as compared to standard electron-phonon systems
\item \gluehg is dominated by a peak (55 meV) in the phonon
    range.
\end{itemize}
The novel idea that follows from these observations is that the
glue function of the cuprates is made up out of electronic and
phononic contributions. Below we will come back to this point
and try to estimate the separate contributions.

\section{Results}\label{results}
\begin{figure}[bht]
\begin{minipage}{17pc}
\includegraphics[width=17pc]{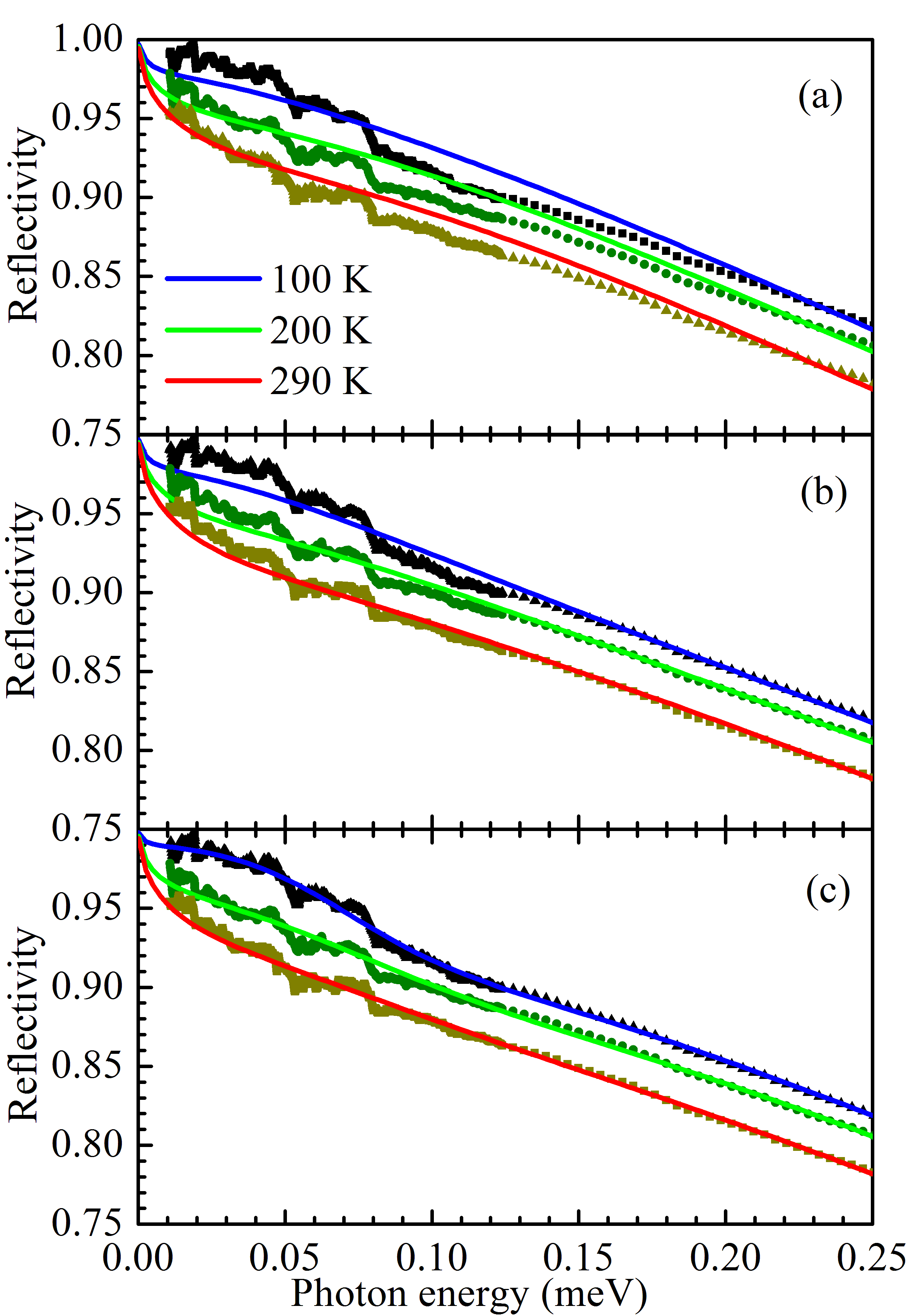}
\caption{\label{refl}Reflectivity of Hg-1201 together with fits for three different models. (a): MFL model, (b): SF model and (c): histogram.}
\end{minipage}\hspace{2pc}%
\begin{minipage}{18pc}
\includegraphics[width=17pc]{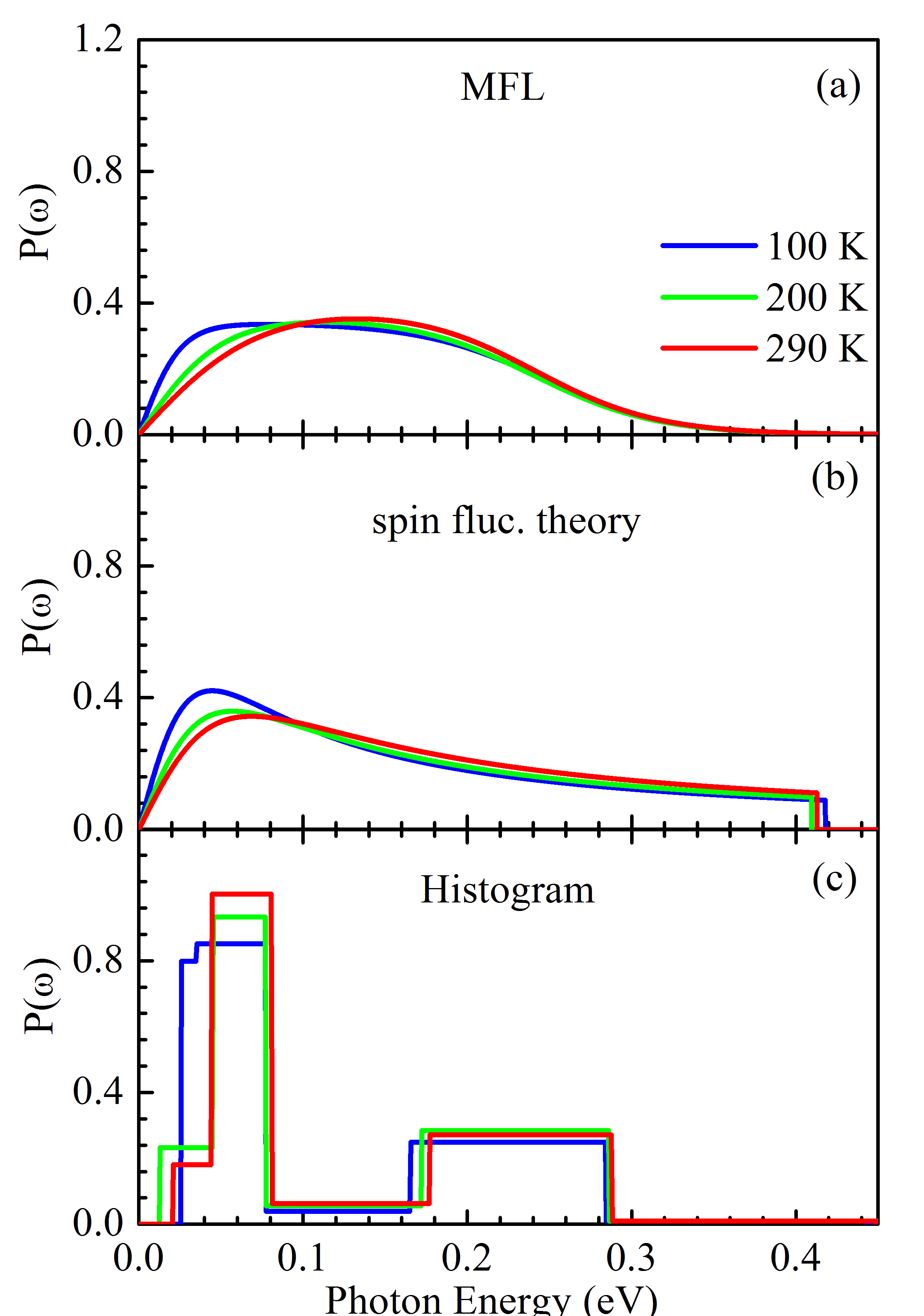}
\caption{\label{a2F}Temperature dependent spectral functions corresponding to the fits shown in the figure \ref{refl}. (a): MFL model, (b): SF model and (c): histogram.}
\end{minipage}
\end{figure}

We analyze optical spectra of optimally doped
HgBa$_{2}$CuO$_{4+\delta}$ (Hg-1201). This high temperature
superconductor presents us with a good opportunity to test the
above models without too many complications. It is a single
layer compound  with a simple tetragonal structure and it has a
high critical temperature of T$_{c}\approx$ 97 K. The optical
data has been presented before in Ref. \cite{heumen-PRB-2007}.
We do not take into account the opening of a (pseudo-)gap in
our analysis so we can only apply it to temperatures of 100 K
and higher. Recent neutron scattering experiments
\cite{li-nat-2008} show that at optimal doping the pseudogap
temperature is probably smaller or equal to the critical
temperature for these materials at optimal doping, although
specific heat measurements on the particular sample used in
this study show an onset of the specific heat jump at slightly
higher temperature \cite{heumen-PRB-2007}. In figure \ref{refl}
we show the reflectivity for Hg-1201 together with the
optimized model curves obtained with Eq.'s (\ref{MFL}),
(\ref{SF}) and (\ref{HG}). The corresponding spectral functions
are shown in figure \ref{a2F}. Figure \ref{refl}a and
\ref{a2F}a show the results for the MFL model. At 290 K we can
make a reasonable fit to the reflectivity but with decreasing
temperature the fit becomes progressively worse. Note that we
try to fit the overall frequency dependence of $R(\omega)$ and
not the sharper phonon structures below 80 meV. It is important
to note that in order to optimize the mean squared deviation,
$\chi^{2}$ (see appendix), we have to make the coupling
constant $\lambda$ in Eq. \ref{MFL} temperature dependent. This
means that already at this level the $\omega/T$ scaling is lost
even for the MFL model. Our result obtained at room temperature
is similar to that obtained in an earlier study
\cite{hwang-PRB-2004}. We find that if we add an extra
scattering channel to the MFL model the fit is greatly improved
(see below).
\glueSF (panels \ref{refl}b and \ref{a2F}b) shows a similar
trend as seen for \glueMFL: a reasonable fit at high
temperature but less good at low temperature. Finally, figures
\ref{refl}c and \ref{a2F}c show \gluehg which gives the best
description of the optical data. The panels of figure \ref{a2F}
give a lot of information on the structure of \glue. As
mentioned above it shows three important features: (i) \glue
has to be made temperature dependent. (ii) there is a high
energy scale that is determined by the cutoff in \glue around
300 meV. (iii) there is a low energy scale determined by a
maximum in \glue around 55 meV. We do not find evidence for
$\omega/T$ scaling of the function \glue from the above three
model calculations. Even in the case of the MFL model we have
to adjust the coupling constant $\lambda$ for each temperature
and this destroys the perfect $\omega/T$ scaling.

To objectively compare the different models we use $\chi^{2}$,
(Eq. \ref{chi}). Table \ref{chisq} lists the optimized
$\chi^{2}$ for the fits shown in figure \ref{refl}. From the
table it follows that overall the SF model performs better than
the MFL model. The reason for this is that the SF model has a
maximum around 55 meV similar to the peak seen in the histogram
representation. The MFL model in contrast has a broad maximum
around 150 meV.
\begin{table}[htb] \caption{\label{chisq}$\chi^{2}$
defined by Eq. \ref{chi} for the three different models. The
values in parentheses are obtained when a Lorentzian peak (Eq.
\ref{Lor}) is added to the spectral function.}
\begin{center}
\begin{tabular}{llll}
\br
Model& 290 K & 200 K & 100 K\\
\mr
MFL&117 (13)&127 (21)&163 (22)\\
SF&23 (15)&37 (21)&74 (18)\\
HG&11&19&13\\
\br
\end{tabular}
\end{center}
\end{table}
The histogram representation suggests that \glue consists of
two contributions: a peak on top of a broader continuum. In
\cite{heumen-sub-2008} we showed that this is true for a whole
range of dopings and temperatures of single, double and triple
layer cuprates. It is therefore tempting to ascribe the peak to
oxygen vibrations present in this energy range while the
continuum indicates coupling to either spin fluctuations or
loop current excitations. There are different interpretations
of our observations possible however.

A first possibility is that one of the assumptions underlying
the strong coupling theory is simply no longer valid. In these
materials the Fermi energy is quite small while the average
mode energy is relatively high, which could mean that the
Migdal approximation, which consists of neglecting vertex
corrections, is no longer valid. This could mean that the high
energy part of the background would be absent if these
corrections are properly taken into account. However, Chubukov
and Schmalian \cite{chubukov-PRB-2005} considered 3D fermions
coupled to a massless bosons spectrum and showed that vertex
corrections can be neglected even in the strong coupling limit.

A second possibility is that we are interpreting parts of the
electronic response in terms of \glue which is actually due to
strong correlation effects. According to Anderson
\cite{anderson-SC-2007} the physics is determined by two energy
scales: the antiferromagnetic exchange coupling $J$ and the
Hubbard repulsion $U$. Since $J\approx$ 0.15 eV it is possible
that we are incorporating features in $\tilde{Pi}(\omega)$
which are actually related to RVB like fluctuations related to
$J$.

The comparison of the panels in figure \ref{a2F} suggests that
we may get a better fit if we add a narrow peak to the MFL or
SF model. As shown in table \ref{chisq} the addition of this
peak dramatically improves the quality of the fit, in
particular for the MFL model. We can now easily separate the
contribution due to the peak and the one due to the continuum
which is not possible for the histogram representation. For
each spectrum we can calculate the coupling constant which is
given by,
\begin{equation}
\lambda=2\int_{0}^{\infty}\frac{\tilde{\Pi}(\omega)}{\omega}d\omega.
\end{equation}
and separate the contribution from the 55 meV peak and the
continuum. The total coupling constant at T = 290 K from the
histogram method is $\lambda_{HG}$ = 1.85 while $\lambda_{MFL}$
= 1.9 and $\lambda_{SF}$ = 1.8. In the latter two cases the
peak contributions are $\lambda_{peak,MFL}$ = 0.88 and
$\lambda_{peak,SF}$ = 0.65 respectively. The peak energy is
somewhat lower in energy than the low energy dispersion kink
seen in ARPES experiments \cite{wslee-condmat-2006}. The
coupling constants obtained by ARPES are of the order
$\lambda\approx$ 0.3 - 0.5
\cite{johnson-PRL-2001,lanzara-nat-2001,cuk-PRL-2004,cuk-PSSB-2004,non-PRL-2006}
as are the coupling constants derived from LDA
\cite{savrasov-PRL-1996,jepsen-JPCS-1998}. The remaining
coupling constant $\lambda_{cont}\approx$ 1.2 for the SF
contribution also corresponds well with earlier estimates
\cite{fink-PRB-2006}. When the temperature is decreased the
total coupling constant increases: $\lambda_{200 K}$ = 2.0 and
$\lambda_{100 K}$ = 2.3. Figure \ref{a2F}c indicates that this
increase arises not simply from an increase in coupling to the
mode but rather from an increase in intensity in the energy
range below 50 meV. Since we cannot exclude pseudogap effects
to play a role here we refrain from separating the
contributions due to the mode and the background.

\section{Conclusion}\label{Revs}
We have presented a detailed analysis of optical spectra of the
high temperature superconductor HgBa$_{2}$CuO$_{4+\delta}$ in
terms of strong coupling theory. A comparison of the marginal
Fermi liquid model and the spin fluctuation model shows that
the latter better describes the optical spectra of Hg-1201.
However, the best description of the optical data is obtained
when a low energy peak is added to either of the two models. A
histogram model of \glue is dominated by a mode at 55 meV and
this is lacking in the previous models. The addition of an
extra mode to the MFL and SF models greatly improves the
agreement with experiment for both models and allows us to
separate the \glue spectrum into two components. The first is a
nearly temperature independent peak centered at 55 meV with a
coupling strength at room temperature $\lambda_{mode}\approx$
0.7. The energy and coupling strength of this mode strongly
suggest an interpretation in terms of electrons coupling to a
vibrational mode. The second feature is a continuum with a
maximum around 200 meV with a coupling strength of about
$\lambda_{cont}\approx$ 1.2.

\section{Acknowledgements}
We would like to acknowledge stimulating discussions with C.M.
Varma, A.V. Chubukov and M.R. Norman, J. Zaanen, D.J. Scalapino
and C. Berthod. This work is supported by the Swiss National
Science Foundation through Grant No. 200020-113293 and the
National Center of Competence in Research (NCCR) "Materials
with Novel Electronic Properties - MaNEP".

\appendix
\section{}\label{App} The calculation of the optical conductivity
based on \glue rests on several assumptions. The most important
are (i) the system is a Fermi liquid with a constant density of
states, (ii) the electrons are coupled to bosonic modes that
have little or no dispersion in k-space and (iii) one can
ignore vertex corrections. For conventional superconductors
these assumptions have been shown to be reasonable but they are
not \textit{apriori} correct when strong correlations effects
are important. Calculations have shown that these assumptions
still apply when the bosons are spin fluctuations with a
spectrum that extends to relatively high energy in the strong
coupling regime \cite{chubukov-PRB-2005}. With the above
approximations the complex optical conductivity
$\hat{\sigma}(\omega)$ can be expressed in terms of a self
energy $\Sigma(\omega)$ in the following way
\cite{pballen-PRB-1971, pballen-book-1982},
\begin{equation}\label{Kubo}
\hat{\sigma}(\omega ,T) = \frac{{\omega _p^2 }}{{i4\pi\omega}}\int\limits_{ -
\infty }^{ + \infty }  \frac{{n_{F}(\omega + x,T) - n_{F}(x,T)dx}}{{\omega  - \Sigma (x + \omega ,T)
+ \Sigma ^* (x,T) + i\Gamma_{imp}}},
\end{equation}
where $\omega_{p}$ is the plasma frequency, $\Gamma_{imp}$ is
an impurity scattering rate, $n_{F}(x)=(exp(\beta x)+1)^{-1}$
is the Fermi-Dirac distribution function and
$\beta=(k_{b}T)^{-1}$. The self energy in Eq. \ref{Kubo} is
given by,
\begin{equation}\label{SE}
 \Sigma \left( {\omega ,T} \right) = \int {d\varepsilon \int d\omega '{\tilde{\Pi}(\omega ')} } \left[\frac{{n_{B} (\omega ') + n_{F} (\varepsilon )}}{{\omega  - \varepsilon  + \omega ' + i\delta }}
  + \frac{{n_{B} (\omega ') + 1 - n_{F} (\varepsilon )}}{{\omega  - \varepsilon  - \omega ' - i\delta }}\right]
\end{equation}
where $n_{B}(x)=(exp(\beta x)-1)^{-1}$ is the Bose-Einstein
distribution function. The integral over $\varepsilon$ can be
performed analytically, leaving,
\begin{equation}\label{SE2}
\Sigma \left( {\omega ,T} \right) = \int d\omega' \tilde{\Pi}(\omega')L\left(\omega,\omega',T\right).
\end{equation}
with
\begin{equation}\label{Lfunc}
L(\omega ,\omega' ,T) =  - i\pi \coth \left( {\frac{\omega
}{{2T}}} \right) + \Psi \left( {\frac{1}{2} + i\frac{{\omega  -
\omega' }}{{2\pi T}}} \right) - \Psi \left( {\frac{1}{2} -
i\frac{{\omega + \omega' }}{{2\pi T}}} \right).
\end{equation}
The $\Psi(x)$ in this last expression are digamma functions.
Equations (\ref{Kubo}-\ref{Lfunc}) together with the standard
Fresnel equations allow us to calculate \textit{any} optical
property given \glue. From these equations one can also
immediately see that it will be very difficult to extract \glue
directly from the experimental data: this would require at
least a second derivative of the optical data
\cite{marsiglio-PLA-1998}. Given this difficulty we use the
following approach. We choose an analytic form for \glue with
several adjustable parameters and numerically perform the
integrations in equations (\ref{Kubo}) and (\ref{SE2}). The
parameters of \glue are optimized using a standard
Levenberg-Marquardt algorithm. This algorithm revolves around
minimizing the function,
\begin{equation}\label{chi}
\chi ^2  \equiv \sum\limits_{i = 1}^N {(\frac{{y_i  - f(x_i ,p_1 ,...,p_m )}}{{\sigma _i }})^{2}}
\end{equation}
where the $y_{i}$ is the experimental datapoint value
corresponding to the point $x_{i}$ and $f(x_{i},p_{1},...,p_{m}
)$ is the calculated value in the point $x_{i}$ from parameters
$p_{1}...p_{m}$ that are to be optimized and the $\sigma_{i}$
are the error bars on the value $y_{i}$. Since the optimization
process requires knowledge of error bars on the quantity to be
fitted, we use the reflectivity spectra measured in the
infrared region of the spectrum for which we have accurate
estimates of the error bars involved. Note that the calculation
of the conductivity requires a choice for the plasma frequency,
$\omega_{p}$. We add $\omega_{p}$ as an independent fit
parameter, but check that it is consistent with values obtained
from a spectral weight analysis. The method described above
applies to the intraband conductivity only. However, our
spectra contain contributions due to interband transitions as
well \cite{heumen-PRB-2007}. These we model using standard
Lorentz oscillators added to the conductivity calculated using
equation (\ref{Kubo}). A second point to be noted is that our
analysis can only be carried out in the normal state. The
analysis of spectra in the superconducting state is more
challenging since it requires solving the full Eliashberg
equations with high precision.\newline

\bibliography{D:/papers/referencefile/reffile}

\providecommand{\newblock}{}
\begin{thebibliography}{10}
\expandafter\ifx\csname url\endcsname\relax
  \def\url#1{{\tt #1}}\fi
\expandafter\ifx\csname urlprefix\endcsname\relax\def\urlprefix{URL }\fi
\providecommand{\eprint}[2][]{\url{#2}}

\bibitem{hussey-nat-2003}
Hussey N~E, Abdel-Jawad M, Carrington A, Mackenzie A~P and Balicas L 2003 {\em
  Nature\/} {\bf 425} 814

\bibitem{takeuchi-PRL-2005}
Takeuchi T, Kondo T, Kitao T, Kaga H, Yang H, Ding H, Kaminski A and Campuzano
  J~C 2005 {\em Phys. Rev. Lett.\/} {\bf 95} 227004 (pages~4)

\bibitem{leyraud-nat-2007}
Doiron-Leyraud N, Proust C, LeBoeuf D, Levallois J, Bonnemaison J~B, Liang R,
  Bonn D~A, Hardy W~N and Taillefer L 2007 {\em Nature\/} {\bf 447} 565--568

\bibitem{basov-science-1999}
Basov D~N, Woods S~I, Katz A~S, Singley E~J, Dynes R~C, Xu M, Hinks D~G, Homes
  C~C and Strongin M 1999 {\em Science\/} {\bf 283} 49--52

\bibitem{molegraaf-science-2002}
Molegraaf H~J~A, Presura C, van~der Marel D, Kes P~H and Li M 2002 {\em
  Science\/} {\bf 295} 2239--2241

\bibitem{syro-EPL-2003}
Santander-Syro A~F, Lobo R~P~S~M, Bontemps N, Konstantinovic Z, Li Z~Z and
  Raffy H 2003 {\em Europhys. Lett.\/} {\bf 62} 568--574

\bibitem{carbone-PRB-2006a}
Carbone F, Kuzmenko A~B, Molegraaf H~J~A, van Heumen E, Lukovac V, Marsiglio F,
  van~der Marel D, Haule K, Kotliar G, Berger H, Courjault S, Kes P~H and Li M
  2006 {\em Phys. Rev. B\/} {\bf 74} 064510 (pages~8)

\bibitem{carbone-PRB-2006b}
Carbone F, Kuzmenko A~B, Molegraaf H~J~A, van Heumen E, Giannini E and van~der
  Marel D 2006 {\em Phys. Rev. B\/} {\bf 74} 024502 (pages~10)

\bibitem{heumen-PRB-2007}
van Heumen E, Lortz R, Kuzmenko A~B, Carbone F, van~der Marel D, Zhao X, Yu G,
  Cho Y, Barisic N, Greven M and Dordevic C~H~S 2007 {\em Phys. Rev. B\/} {\bf
  75} 054522 (pages~10)

\bibitem{kuzmenko-PRB-2005}
Kuzmenko A~B, Molegraaf H~J~A, Carbone F and van~der Marel D 2005 {\em Phys.
  Rev. B\/} {\bf 72} 144503 (pages~9)

\bibitem{varma-PRL-2007}
Aji V and Varma C~M 2007 {\em Phys. Rev. Lett.\/} {\bf 99} 067003 (pages~4)

\bibitem{dirk-nat-2003}
van~der Marel D, Molegraaf H~J~A, Zaanen J, Nussinov Z, Carbone F, Damascelli
  A, Eisaki H, Greven M, Kes P~H and Li M 2003 {\em Nature\/} {\bf 425}
  271--274

\bibitem{norman-PRB-2006}
Norman M~R and Chubukov A~V 2006 {\em Phys. Rev. B\/} {\bf 73} 140501 (pages~4)

\bibitem{norman-PRB-2007}
Norman M~R, Chubukov A~V, van Heumen E, Kuzmenko A~B and van~der Marel D 2007
  {\em Phys. Rev. B\/} {\bf 76} 220509 (pages~4)

\bibitem{marsiglio-PLA-1998}
Marsiglio F, Startseva T and Carbotte J 1998 {\em Phys. Lett. A\/} {\bf 245}
  172

\bibitem{dordevic-PRB-2005}
Dordevic S~V, Homes C~C, Tu J~J, Valla T, Strongin M, Johnson P~D, Gu G~D and
  Basov D~N 2005 {\em Phys. Rev. B\/} {\bf 71} 104529 (pages~12)

\bibitem{varma-PRL-1989}
Varma C~M, Littlewood P~B, Schmitt-Rink S, Abrahams E and Ruckenstein A~E 1989
  {\em Phys. Rev. Lett.\/} {\bf 63} 1996--1999

\bibitem{millis-PRB-1990}
Millis A~J, Monien H and Pines D 1990 {\em Phys. Rev. B\/} {\bf 42} 167--178

\bibitem{greven-condmat-2008}
Li Y, Balédent V, Barisic N, Cho Y, Fauqué B, Sidis Y, Yu G, Zhao X, Bourges P
  and Greven M {\em cond-mat:0805.2959\/}

\bibitem{chubukov-PRB-2005}
Chubukov A~V and Schmalian J 2005 {\em Phys. Rev. B\/} {\bf 72} 174520
  (pages~14)

\bibitem{anderson-SC-2007}
Anderson P~W 2007 {\em Science\/} {\bf 316} 1705--1707

\bibitem{lanzara-nat-2001}
Lanzara A, Bogdanov P~V, Zhou X~J, Kellar S~A, Feng D~L, Lu E~D, Yoshida T,
  Eisaki H, Fujimori A, Kishio K, Shimoyama J~I, Noda T, Uchida S, Hussain Z
  and Shen Z~X 2001 {\em Nature\/} {\bf 412} 510

\bibitem{non-PRL-2006}
Meevasana W, Ingle N~J~C, Lu D~H, Shi J~R, Baumberger F, Shen K~M, Lee W~S, Cuk
  T, Eisaki H, Devereaux T~P, Nagaosa N, Zaanen J and Shen Z~X 2006 {\em Phys.
  Rev. Lett.\/} {\bf 96} 157003 (pages~4)

\bibitem{pballen-PRB-1971}
Allen P~B 1971 {\em Phys. Rev. B\/} {\bf 3} 305--320

\bibitem{pballen-book-1982}
Allen P~B and Mitrovic B 1982 {\em Solid state physics\/} Advances in research
  and applications ed Ehrenreich H, Seitz F and Turnbull D (London: Academic
  Press) p~2

\end{thebibliography}

\end{document}